\documentclass[a4paper]{PoS}

\usepackage{tikz}
\usetikzlibrary{decorations.markings, arrows}

\usepackage{mathtools}
\usepackage{subfig}
\usepackage[capitalize]{cleveref}
\usepackage{microtype}

\title{Worldline Approach to Few-body Physics on the Lattice}

\ShortTitle{Worldline Approach to Few-body Physics on the Lattice}

\author{\speaker{Hersh Singh}\thanks{In collaboration with Shailesh Chandrasekharan, \emph{Duke University}, \texttt{sch@phy.duke.edu}.}\\
        Duke University\\
        E-mail: \email{hersh@phy.duke.edu}}

\newcommand\<{\langle}
\renewcommand\>{\rangle}

\DeclareMathOperator{\Trace}{Tr}
\DeclareMathOperator{\Tr}{Tr}
\newcommand\MCConfig{C}

\newcommand\mbar{\overline{m}}
\newcommand\EnergyFG{E_{\text{FG}}}
\newcommand\EnergyFermi{\epsilon_{\text{F}}}
\newcommand\LX{L_X}
\newcommand\LT{L_T}
\newcommand\Hopping[1]{t_{#1}}
\newcommand\Mass[1]{m_{#1}}

\newcommand\NN{\NumParticles{\uparrow}, \NumParticles{\downarrow}}

\newcommand\EnergyGround{E^{0}_{\NumParticles{\uparrow}, \NumParticles{\downarrow}}}

\newcommand\NumHops[1]{{N_h^{#1}}}
\newcommand\NumParticles[1]{{N_{#1}}}
\newcommand\NumInteractions{{N_I}}
\newcommand\WeightMove{W_m}
\newcommand\WeightHop{W_h}
\newcommand\WeightInt{W_I}

\newcommand\PlotWidthFactor{0.49}

\abstract{
  We study the physics of two species of non-relativistic hard-core bosons with attractive or repulsive delta function interactions on a spacetime lattice using the worldline formulation. By tuning the chemical potential carefully we show that worm algorithms can efficiently sample the worldline configurations in any fixed particle-number sector. Since fermions can be treated as hard-core bosons up to a permutation sign, we also apply this approach to non-relativistic fermions. The fermion permutation sign is treated as an observable in this approach and can be used to extract energies for each particle-number sector. Since in one dimension non-relativistic fermions can only permute due to boundary effects, unlike the auxiliary field method, in many cases our approach does not suffer from sign problems. Using our method we discover limitations of the recently proposed complex Langevin calculations in one dimension.
}

\FullConference{The 36th Annual International Symposium on Lattice Field Theory - LATTICE2018\\
		22-28 July, 2018\\
		Michigan State University, East Lansing, Michigan, USA.}

\begin{document}

\section{Introduction}
Effective field theories of strongly coupled particles with short-range contact interactions are ubiquitous.  For example, in nuclear physics, pionless effective field theory \cite{bedaque_effective_2002} has been successfully used to describe low atomic number nuclei.
In condensed matter physics, this has been applied to cold atoms to explain universal behavior of systems with large scattering lengths \cite{braaten_universality_2006}.
Such theories are also being explored to describe dark matter candidates such as axions \cite{braaten_nonrelativistic_2016} and to look at signatures of bound states of dark matter \cite{laha_direct_2014, braaten_universal_2013}.
In particle physics, EFTs with contact interactions have been used to describe the quarkonium spectra, e.g., exotics such as X(3872) \cite{fleming_pion_2007}.

Analytic computations for such EFTs are very limited in the non-perturbative regime. 
For few-nucleon physics in pionless EFT, for instance, one can do analytic calculations for two-body systems (such as deuteron) and even three-body systems (such as helium-3, triton) to some extent.  However, the problem for anything more than three bodies quickly becomes intractable, except in cases with extra symmetries \cite{bansal_pion-less_2018, contessi_ground-state_2017}.
Further, the renormalization group behavior is still being debated \cite{epelbaum_wilsonian_2017} and will be important for higher-order calculations. 
In the case of many body physics, the unitary Fermi gas offers an example of a non-perturbative system of great theoretical and experimental interest \cite{zwerger_bcs-bec_2011}.

Lattice formulations offer a window into the non-perturbative features of such theories.
The state of the art methods are typically based on auxiliary field Monte Carlo techniques (see recent reviews \cite{lee_lattice_2017, nicholson_lattice_2017} for lattice nuclear EFTs), which unfortunately suffer from sign problems in many cases of interest.
This is even true for systems in one spatial dimension, which have recently become exciting due to the advent of experimental realizations with ultracold atoms in confining traps \cite{guan_fermi_2013}.
One spatial dimension is then a good testing ground for new approaches to solve sign problems, especially since exact solutions are available in several cases \cite{gaudin_systeme_1967, yang_exact_1967, lieb_absence_1968}.
Ref.~\cite{rammelmuller_surmounting_2017} recently used the complex Langevin (CL) approach to study two species of mass-imbalanced fermions in one spatial dimension with a delta-function interaction, given by the continuum Hamiltonian
\begin{align}
  H =  \int\! dx\! \left( -\sum_{\mathclap{\sigma= \uparrow, \downarrow}}\ \ \frac{1}{2m_\sigma} \psi^\dagger_\sigma\,\partial_x^{2} \psi_\sigma + g \psi^\dagger_\uparrow \psi_\uparrow\, \psi^\dagger_\downarrow \psi_\downarrow\right).
  \label{eq:hamiltonian-gaudin-yang}
\end{align}
In one spatial dimension, the sign problem is also absent in the worldline formulation \cite{wiese_bosonization_1993, evertz_loop_2003} under certain conditions. This is because fermions effectively become hard-core bosons up to boundary effects.  Motivated by this, we construct a general worldline approach to study the physics of hard-core bosons, which we describe in the next section.

\section{Model and Parameters}
\label{sec:model}

We work with the following discretization of the Hamiltonian \eqref{eq:hamiltonian-gaudin-yang} for hard-core bosons, 
\begin{align}
  H &= -\sum_{\mathclap{{i, \sigma}}} \ t^{\sigma}\ (c_{i,\sigma}^{\dagger} c_{i+1, \sigma}^{\phantom{\dagger}} + c^\dagger_{i+1, \sigma} c_{i,\sigma}^{\phantom{\dagger}} - 2 c^\dagger_{i,\sigma} c_{i,\sigma}^{\phantom{\dagger}})   
    + \frac{U}{a} \ \sum_i \ \NumParticles{i,\uparrow} \NumParticles{i,\downarrow},
      \label{eq:lattice-model}
\end{align}
where
$c_{i,\sigma}^\dagger, c_{i,\sigma}^{\phantom{\dagger}}$ are bosonic creation and annihilation operators for the two species of particles, say spin up ($\sigma = \uparrow$) and spin down ($\sigma = \downarrow$), at the site $i$ on a one-dimensional spatial lattice,
$a$ is the lattice spacing in a box of size $\LX$, so that the physical box size is $L = \LX a$.
The hopping parameter $\Hopping{\sigma}$ for the species $\sigma$ is related to the mass $\Mass{\sigma}$ of the $\sigma$-particles by $\Hopping{\sigma} = 1/(2\Mass{\sigma} a^2)$.
The two particle species have an on-site interaction term written in terms of site occupation number operators $N_{i, \sigma} = c^{\dagger}_{i,\sigma} c_{i, \sigma}^{\phantom{\dagger}}$ and the bare coupling constant $U$.
In addition, we impose the \emph{hard-core} boson constraint, which means that we do not allow states with more than one particle of a given type on each site; in other words, the site occupation numbers for each species can only take two values: $\NumParticles{i,\sigma} \in \{0, 1\}$.  Conveniently, there is no subtlety in taking the continuum limit in this case. The lattice model \eqref{eq:lattice-model} reproduces the continuum Hamiltonian \eqref{eq:hamiltonian-gaudin-yang} in the naive continuum limit $a\to 0$ with $U=g$.

Let $\NumParticles{\sigma} = \sum_{i} N_{i, \sigma} $ be the number of $\sigma$-particles, so that $N = \NumParticles{\uparrow} + \NumParticles{\downarrow}$ is the total number of particles, and let $n=N/L$ be the number density.  
We then parameterize the interaction strength by $\gamma = U/n$.  To discuss mass-imbalanced systems, we use the mass-imbalance parameter $ \mbar = \frac{m_\uparrow - m_\downarrow}{m_\uparrow + m_\uparrow} $ and the average mass $ m = \frac{m_\uparrow + m_\downarrow}{2} $.
We report all energies in units of the Fermi gas energy $\EnergyFG = N \EnergyFermi/3$, which is the ground-state energy of spin- and mass-balanced $N$ free fermions in the continuum, with
$\EnergyFermi = \pi^2n^2/8m$ being the Fermi energy of this system.  We also set average mass $m=1$ and lattice spacing $a=1$ throughout.

In this work, we shall focus on computing the ground-state energy $\EnergyGround$ of the lattice model \eqref{eq:lattice-model} for hard-core bosons in the $(\NumParticles{\uparrow}, \NumParticles{\downarrow})$ particle-number sector in one spatial dimension.  To develop a worldline approach for computing the ground-state energy, we define the partition function in a given particle-number subspace,
\begin{align}
  Z_\mu^{\NN}(\beta) = \Tr\left(e^{-\beta H_\mu}\right)\Big|_{{\NN}},
  \label{eq:partition-function-mu}
\end{align}
where $\beta$ is the inverse temperature, and we use the Hamiltonian defined in \cref{eq:lattice-model} with a chemical potential $\mu_\sigma$ for the species $\sigma$,
\begin{align}
  H_\mu = H - \sum_{\mathclap{\sigma = \{\uparrow, \downarrow\}}}\ \mu_\sigma N_\sigma.
\end{align}
The average energy at a given $\beta$ is then,
\begin{align}
  \< E \>_{\NN} = \frac{1}{Z_\mu^{\NN}} \Trace\left.\left( H_\mu e^{-\beta H_\mu} \right) \right|_{\NN}.
  \label{eq:energy-observable-mu}
\end{align}
Due to the restriction to a fixed particle-number subspace, the chemical potential drops out of this expression and we indeed get the ground-state energy of $H$ in the $(\NN)$ sector for large $\beta$,
\begin{align}
  \EnergyGround = \lim_{\beta \to \infty} \< E \>_{\NN}.
\end{align}
If we can construct an efficient method to sample from the partition function \eqref{eq:partition-function-mu}, then we can compute the ground-state energies easily.  This can be achieved using worm algorithms in the worldline formulation with a careful tuning of the chemical potential.

\section{Worldline Formulation and the Worm Algorithm}
\label{sec:worldline-formulation}

We write the Hamiltonian as $H = H_{h} + H_{d}$, where $H_{d}$ is diagonal in the occupation number basis, and $H_{h}$ is the off-diagonal hopping term.  Expanding the partition function as a Dyson series, we get
\begin{align}
  Z_\mu = &\sum_k\ \int_0^\beta\!\! dt_k \int_0^{t_k}\!\!  dt_{k-1} \cdots \int_0^{t_2}\!\!  dt_1 
        \Tr\Bigg(
          e^{-(\beta-t_k) H_d} H_h
          e^{-(t_k-t_{k-1}) H_d}H_h \cdots H_h 
          e^{-t_1 H_d}\Bigg).
          \label{eq:Z-dyson-series}
\end{align}
Even though this can be done in continuous time, it is convenient at this point to divide $\beta$ into $\LT$ time steps with spacing $\varepsilon$, so that $\LT \varepsilon = \beta$.
Computing this trace in the occupation number basis lets us write $ Z_\mu = \sum_{\MCConfig} \Omega(\MCConfig) $,
where $\MCConfig$ is a worldline configuration on a $\LX \times \LT$ spacetime lattice and $\Omega(\MCConfig)$ is the weight of the configuration $\MCConfig$. The weight can be computed from the total number of spatial hops $\NumHops{\sigma}(\MCConfig)$, number of particles $\NumParticles{\sigma}(\MCConfig)$ and the number of interactions $\NumInteractions(\MCConfig)$ using the formula
\begin{align}
  \Omega(\MCConfig) &= 
                 (\WeightMove^\uparrow)^\NumHops{\uparrow}
                 (\WeightMove^\downarrow)^\NumHops{\downarrow}
                 (\WeightHop^\downarrow)^\NumParticles{\downarrow}
                 (\WeightHop^\uparrow)^\NumParticles{\uparrow}
                 (\WeightInt)^\NumInteractions.
\end{align}
where $\WeightMove^{\sigma}$, $\WeightHop^{\sigma}$ and $\WeightInt$ are local weights arising from each spatial hop, temporal move (forward in time), and interactions, respectively:
\begin{align}
W^{\sigma}_h = \Hopping{\sigma}\varepsilon, \quad 
W^{\sigma}_m = \exp\big(-\varepsilon (2 \Hopping{\sigma}- \mu_\sigma)\big),\quad 
\WeightInt = \exp\left(-\varepsilon U\right).
\end{align}
\Cref{fig:worldline-example} shows an example of such a configuration.   
The weight of the configuration is the product of all the local site weights.  For each such configuration $C$, we can define its energy $E(C)$ as
\begin{align}
E(\MCConfig) &= \frac{U \NumInteractions(\MCConfig) }{L_T} +  \sum_{\sigma} \left[ - \frac{\NumHops{\sigma}(\MCConfig) }{{\beta}} + 2 \Hopping{\sigma} \NumParticles{\sigma}({\cal C}) \right].
\label{eq:energy-of-worldline}
\end{align}
The average energy is then 
\begin{align}
  \langle E \rangle &=  \frac{1}{Z_\mu} \sum_{\MCConfig} E(\MCConfig) \ \Omega(\MCConfig).
\label{eq:energy-average}
\end{align}
which is the same as the average energy defined in \cref{eq:energy-observable-mu} in the $\epsilon\to 0$ limit.

\begin{figure}[htb]
  \centering
  \begin{minipage}{0.45\linewidth}
    \newcommand\FigWeightMove[1]{W^{#1}_m}
\newcommand\FigWeightHop[1]{W^{#1}_h}
\newcommand\FigWeightInteraction{W_I}

\tikzset{
    dots/.style args={#1per #2}{%
      line cap=round,
      dash pattern=on 0 off #2/#1
    },
    inner grid/.style = {ultra thin, step=1.0, opacity=0.3},
    outer grid/.style = {line width=0.4mm, step=1.0, black, dots=15 per 1cm, opacity=0.5},
    site/.style = {circle, inner sep=0mm, minimum size=1.5mm, fill=black, opacity=0.4},
}

\def\InteractionColor{green!50!black}

  \begin{tikzpicture}[
    mid arrow/.style={postaction={decorate,decoration={
          markings,
          mark=at position .6 with {\arrow[#1]{stealth}}
        }}},
    weight label down/.style = {right,},
    weight label up/.style = {left,},
    node distance=0.5mm,
    label distance=0.5mm,
    scale=1.0,
    worldline blue/.style = {ultra thick, color=blue!70!black, opacity=1.0},
    worldline red/.style = {ultra thick, color=red!70!black, opacity=1.0},
    worldline interaction/.style = {\InteractionColor, ultra thick },
    ]

    \begin{scope}
      \clip(-2,0) rectangle (4,5);
      \draw[xshift=0.5cm, yshift=0.5cm, inner grid] (-2.5,-0.5) grid (4,5);
    \end{scope}
    
    \draw[outer grid] (-2,0) grid (4,5);

    \begin{scope}
      \clip(-2,0) rectangle (4,5);

      \draw [worldline blue]
      (0.5,-0.5) edge[mid arrow] (0.5,0.5) (0.5,0.5) node [weight label up, right] {$\FigWeightMove{\uparrow}$} 
      coordinate[] edge[mid arrow] ++(0,1) ++(0,1)  node [weight label up, right] {$\FigWeightHop{\uparrow}$} 
      coordinate[] edge[mid arrow] ++(-1,0)++(-1,0) node [weight label up, ] {$\FigWeightMove{\uparrow}$} 
      coordinate[] edge[mid arrow] ++(0,1) ++(0,1) node [weight label up] {$\FigWeightHop{\uparrow}$} 
      coordinate[] edge[mid arrow] ++(1,0) ++(1,0) node [weight label up, above] {$\FigWeightHop{\uparrow}$} 
      coordinate[] edge[mid arrow] ++(1,0) ++(1,0) node [weight label up, ] {}
      coordinate[] edge[mid arrow] ++(0,1)  ++(0,1)  node [weight label up, above left] {$\FigWeightHop{\uparrow}$} 
      coordinate[] edge[mid arrow] ++(-1,0) ++(-1,0) node [weight label up] {$\FigWeightMove{\uparrow}$} 
      coordinate[] edge[mid arrow] ++(0,1)  ++(0,1)  node [weight label up] {$\FigWeightMove{\uparrow}$} 
      coordinate[] edge[mid arrow] ++(0,1)  ++(0,1)  node [weight label up] {$\FigWeightMove{\uparrow}$} ;


      \draw [worldline red]
      (2.5, -0.5) edge[mid arrow] (2.5,0.5) (2.5, 0.5) node [weight label down] {$\FigWeightMove{\downarrow}$}
      coordinate [] edge[mid arrow] ++(0,1) ++(0,1) node [weight label down] {$\FigWeightHop{\downarrow}$}
      coordinate [] edge[mid arrow] ++(0,1) ++(0,1) node [weight label down] {$\FigWeightHop{\downarrow}$}
      coordinate [] edge[mid arrow] ++(-1,0) ++(-1,0) node [weight label down, below, \InteractionColor] {$\FigWeightMove{\uparrow} \FigWeightMove{\downarrow} \FigWeightInteraction$} 
      coordinate [] edge[mid arrow, draw=white] ++(0,1) ++(0,1) node [weight label down, above right] {$\FigWeightHop{\downarrow}$}
      coordinate [] edge[mid arrow] ++(1,0) ++(1,0) node [weight label down] {$\FigWeightMove{\downarrow}$}
      coordinate [] edge[mid arrow] ++(0,1) ++(0,1) node [weight label down] {$\FigWeightMove{\downarrow}$}
      coordinate [] edge[mid arrow] ++(0,1) ++(0,1) node [weight label down] {$\FigWeightMove{\downarrow}$};

      \draw [worldline interaction] (1.5,2.5) edge[mid arrow] ++(0,1);
    \end{scope}
    
    \foreach \x in {-1.5,...,3.5} {
      \foreach \y in {0.5,...,4.5} {
        \node [site] (N-\x-\y) at (\x,\y) {};
      }
    }

  \end{tikzpicture}

    \centering
    \caption{An example worldline configuration with two species of particles, with the local weights shown at each site.  The horizontal axis is space and the vertical axis is time.  The interaction of two particles is shown with a differently colored bond. The weight of the configuration is the product of all the site weights.}
    \label{fig:worldline-example}
  \end{minipage}\hfill
  \begin{minipage}{0.45\linewidth}
    \centering
    \def\FigureScale{0.55}
\tikzset{
  boundary line/.style = {
    thick
  },
  space boundary line/.style = {
    thick, opacity=0.5, dashed
  },
  worldline/.style = {
    thick, 
    decoration={
      markings,
      mark=at position 0.5 with {\arrow{stealth}}},
    postaction={decorate},
  }, 
  particle node/.style= {
    circle, draw, fill=black, opacity=0.8,
    inner sep=0pt,minimum size=5pt
  }
}
\begin{tikzpicture}[scale=\FigureScale]
  \def\xLen{4.5}
  \def\yLen{5}
  \def\numParticles{4}
  \node [above] at (\xLen/2, \yLen) {$\text{Sign} = +1$};

  \draw [boundary line] (0.5,0) -- (\xLen,0);
  \draw [boundary line]  (0.5,\yLen) -- (\xLen,\yLen);
  \draw [space boundary line] (0.5,0) -- (0.5, \yLen);
  \draw [space boundary line] (\xLen,0) -- (\xLen, \yLen);

  \def\axisYShift{1}
  \def\axisXShift{0.5}
  \draw [yshift = -1cm, -stealth] (-\axisXShift,0) -- (\xLen/2,0) node [right] {$x$};
  \draw [xshift = -0.5cm, -stealth] (0,-\axisYShift) -- (0,\yLen) node [above] {$t$};

  \foreach \i in {1,...,\numParticles}{
    \node[particle node] at (\i,0) {};
    \node[particle node] at (\i,5) {};
  }
  \foreach \i in {1,...,\numParticles}{
    \draw [worldline] (\i,0) to[out=90, in=60+180] (\i+0.1, 2.5);
    \draw [worldline] (\i+0.1, 2.5) to[out=60, in=270] (\i, 5);
  }
\end{tikzpicture}~~~~~~
\begin{tikzpicture}[scale=\FigureScale]
  \def\xLen{4.5}
  \def\yLen{5}
  \def\numParticles{4}
  \node [above] at (\xLen/2, \yLen) {$\text{Sign} =-1$};

  \draw [boundary line] (0.5,0) -- (\xLen,0);
  \draw [boundary line]  (0.5,\yLen) -- (\xLen,\yLen);
  \draw [space boundary line] (0.5,0) -- (0.5, \yLen);
  \draw [space boundary line] (\xLen,0) -- (\xLen, \yLen);
  
  \foreach \i in {1,...,\numParticles}{
    \node[particle node] at (\i,0) {};
    \node[particle node] at (\i,5) {};
  }
  \foreach \i in {1,...,\numParticles}{
    \draw [worldline] (\i,0) to[out=90, in=60+180] (\i+0.5, 2.5);
    \ifnum \i = \numParticles
    \else \draw [worldline] (\i+0.5, 2.5) to[out=60, in=270] (\i+1, 5);
    \fi
  }
  \def\i{\numParticles}
  \draw [worldline] (\i,0) to[out=90, in=60+180] (\i+0.5, 2.5);
  \def\i{0} \draw [worldline] (\i+0.5, 2.5) to[out=60, in=270] (\i+1, 5);

  \draw [
  thick,
  opacity=0.25,
    decoration={
      markings,
      mark=at position 0.55 with {\arrow{stealth}}},
    postaction={decorate},
  ] (4.5,2.5) to[out=30, in=150] (\i+0.5, 2.5);

  \begin{scope}[]

    \def\axisYShift{1}
    \def\axisXShift{0.5}
    \draw [yshift = -1cm, -stealth] (-\axisXShift,0) -- (\xLen/2,0) node [right] {$x$};
    \draw [xshift = -0.5cm, -stealth] (0,-\axisYShift) -- (0,\yLen) node [above] {$t$};
  \end{scope}
\end{tikzpicture}
 \\
    \caption{An illustration of how the fermionic sign arises in the worldline formulation.  The configuration on the left has no permutation of the worldline configurations and hence no sign.  In the right figure, one of the particle worldlines wraps around the edge of the periodic box causing all the fermions to shift to the right by one position. This is an odd permutation, which gives a negative sign.}
    \label{fig:fermionic-sign}
  \end{minipage}
\end{figure}

\subsection{Fermions and the Sign Problem in 1+1D}
\label{sec:fermion-sign-problems}

Our model is defined with hard-core bosons.  If we have fermions instead, the trace in \cref{eq:Z-dyson-series}, when evaluated in the occupation number basis, would still lead to a sum over the same worldline configurations but with an additional sign from the permutation of fermionic worldlines.  That motivates the definition of the \emph{fermionic} partition function as $ Z_f = \sum_{\MCConfig} S(\MCConfig) \Omega(\MCConfig) $, 
where $S(C)$ is the sign associated with the configuration $C$. 
We can now see how the worldline approach leads to a solution of the sign problem \cite{wiese_bosonization_1993}.
In one spatial dimension with only nearest neighbor hops, the particle worldlines cannot cross each other.  So the sign can come only from a cyclic permutation, as shown in \Cref{fig:fermionic-sign}.  For fermions in a box with periodic boundary conditions, the sign of a configuration from such a cyclic permutation is
\begin{align}
  S(\MCConfig) = (-1)^{\sum_{\sigma} (\NumParticles{\sigma}(\MCConfig) - 1)N_{w}^{\sigma}(\MCConfig)},
\end{align}
where $N_{w}^{\sigma}$ is the number of times the $\sigma$-particle worldlines wind around the periodic box.  In particular, if $N_\uparrow$ and $N_\downarrow$ are both odd, then $S(\MCConfig)=1$.  Moreover, with open boundary conditions or in a trapping potential, the fermions cannot wind around the box, so $N_{w}^{\sigma}=0$ and again the sign problem is solved.  As promised, fermions become hard-core bosons in all these cases.  However, this argument does not work in higher dimensions since fermions can simply wind around each other, and the sign problem returns.

\section{Results for Fermions in One Spatial Dimension}
\label{sec:results}

\begin{figure*}[htb]
  \centering
  \subfloat[\label{fig:55-mbar-0.0}]{
    \includegraphics[width=\PlotWidthFactor\linewidth]{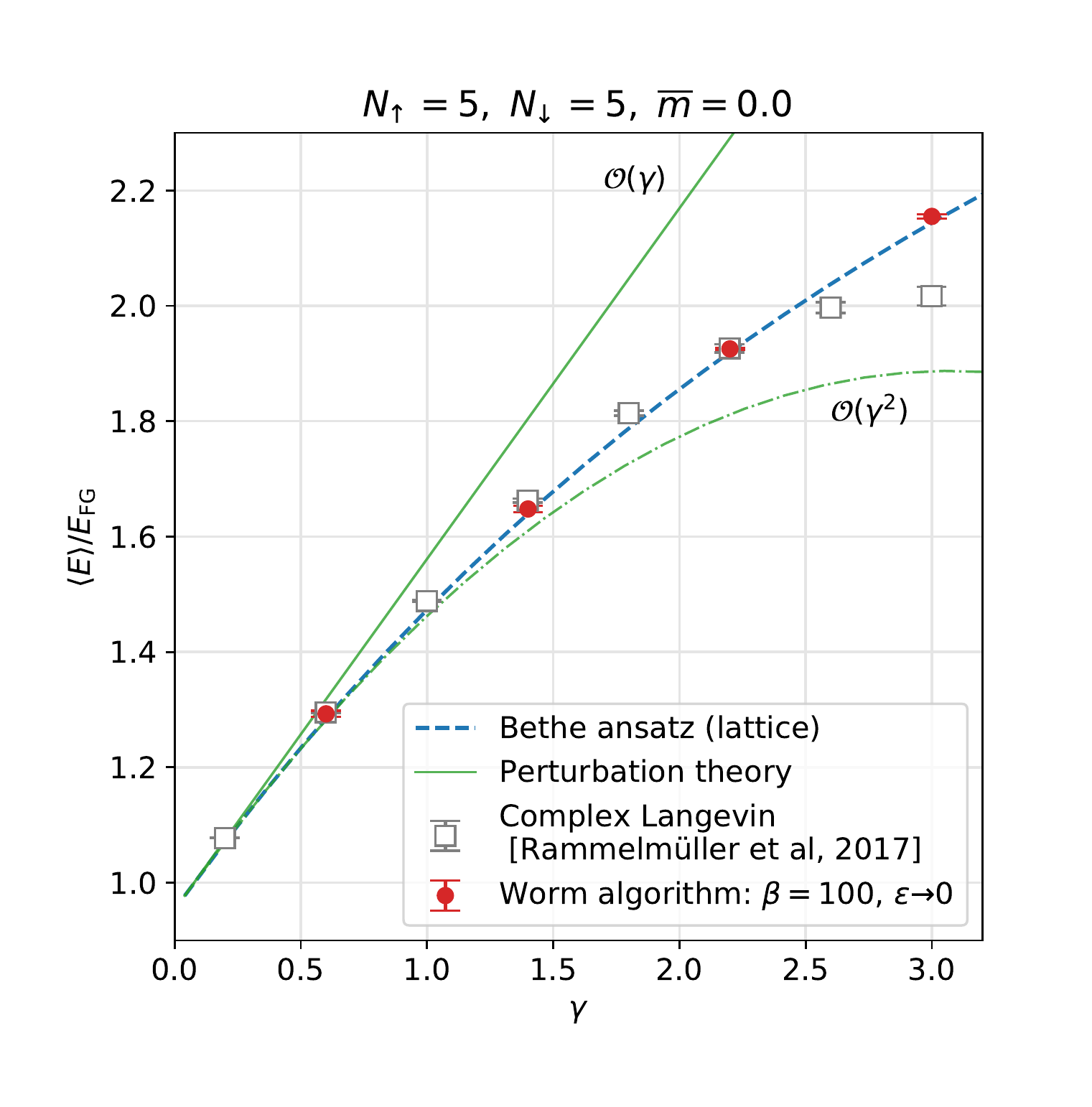}
  }
  \subfloat[\label{fig:55-mbar-0.8}]{
    \includegraphics[width=\PlotWidthFactor\linewidth]{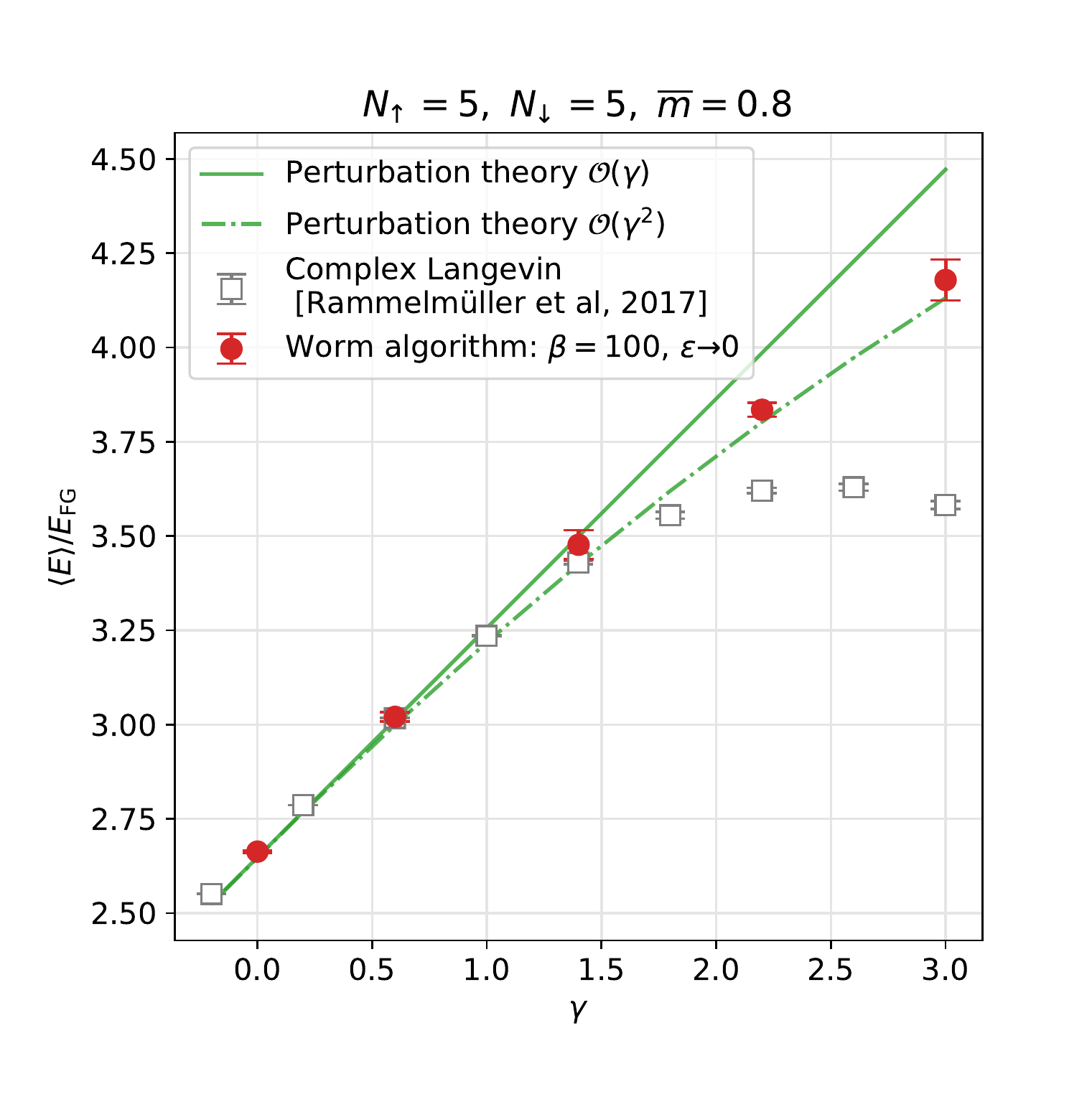}
  }
  \caption{ Plots of the ground-state energy for $N_\uparrow=N_\downarrow=5$ fermions, in a box of size $\LX=40$, as a function of coupling strength $\gamma$ in the repulsive regime for the mass-balanced case $\mbar=0.0$ (left) and high mass imbalanced case $\mbar=0.8$ (right).  For $\mbar=0.0$, we also show the exact results computed from the Bethe ansatz.  The CL results from Ref.~\cite{rammelmuller_surmounting_2017} begin to deviate from the exact results and the worm algorithm around $\gamma \gtrsim 2$. For the mass-imbalanced case (right), we find disagreement even in the regime where second-order perturbation theory works well. }
  \label{fig:55-results}
\end{figure*}

As a check of our method, we compute the ground-state energy for the mass-balanced case, that is, $\mbar=0$.
For fermions, this is the Hubbard model and can be exactly solved using the Bethe ansatz \cite{lieb_absence_1968}.
\Cref{fig:55-mbar-0.0} shows our results for the ground-state energy of $N_\uparrow = N_\downarrow = 5$ fermions with a varying repulsive interaction in a box of size $\LX=40$.  Our results are in excellent agreement with the exact results.  We also compare with the CL results published in Ref.~\cite{rammelmuller_surmounting_2017}.
We find the CL results to be in disagreement with both the Bethe ansatz and the worm algorithm for moderately strong repulsive couplings.

We then consider the case of mass-imbalanced fermions with a repulsive interaction.
\Cref{fig:55-mbar-0.8} shows our results for $N_\uparrow = N_\downarrow = 5$ fermions with a high mass-imbalance $\mbar=0.8$ in a box of size $\LX=40$.
In this range of coupling, we find excellent agreement with second-order perturbation theory. 
However, we find the CL results from Ref.~\cite{rammelmuller_surmounting_2017} to again be in disagreement with our results.  The authors of Ref.~\cite{rammelmuller_surmounting_2017} note a `flattening' of the ground-state energy for strong repulsive couplings, which we do not see with our method.
However, they do point out that the observables in the CL approach have `fat-tailed distributions' for repulsive couplings, which may be a hint that CL is converging to incorrect results in this regime.  As we see in \cref{fig:55-results}, this is indeed the case.

\section{Conclusions}
We developed a worldline approach to study the few-body physics of two species of hard-core bosons.  In one spatial dimension, fermions become hard-core bosons in this approach and we can study fermions without a sign problem.
We showed that we can efficiently compute ground-state energies in any given particle-number sector using the worm algorithm with a carefully tuned chemical potential.
We found that the CL method of Ref.~\cite{rammelmuller_surmounting_2017} converges to the wrong results for repulsive interactions as the coupling strength grows, as suspected by the authors.  We verified this using exact computations from the Bethe ansatz for the mass-balanced case, and second-order perturbation theory.
Our results here pave the way for further studies of more general contact EFTs on the lattice in higher dimensions, and provide an important benchmark for other methods, such as complex Langevin, being used to circumvent the sign problem.

\section*{Acknowledgments}
We would like to thank Lukas Rammelmüller, Jens Braun, Joaquin Drut for useful conversations and for sharing their data published in Ref.~\cite{rammelmuller_surmounting_2017} with us. 
We would also like to thank Roxanne Springer, Dean Lee and Abhishek Mohapatra for helpful discussions. 
This work was supported by the U.S. Department of Energy, Office of Science, Nuclear Physics program under Award No. DE-FG02-05ER41368.


\providecommand{\href}[2]{#2}\begingroup\raggedright\endgroup

\end{document}